\documentclass[apj]{emulateapj-rtx4} 
\usepackage{amsmath}
\usepackage{natbib}
\usepackage{lscape}

\shorttitle{Cold Debris Disk Temperature and Stellar Type}
\shortauthors{Ballering et al.}
\slugcomment{Accepted for publication in ApJ}

\newcommand{\NumTargs}{546}
\newcommand{\NumNoExcess}{321}
\newcommand{\NumExcess}{225}
\newcommand{\NumOneAndCold}{100}
\newcommand{\NumOneAndWarm}{51}
\newcommand{\NumTwoComps}{74}
\newcommand{\NumColdComps}{174}
\newcommand{\NumUpperLims}{25}
\newcommand{\Slope}{0.00568}
\newcommand{\SlopeSigma}{0.00119}

\newcommand{\Intercept}{24.0}

\newcommand{\Scatter}{19.0}
\newcommand{\ScatterSigma}{1.46}
\newcommand{\ColdsBeyondHund}{29}
\newcommand{\ColdsBeyondHundTwenty}{19}
\newcommand{\ColdsBeyondHundFifty}{4}

\begin{document}

\title{A Trend Between Cold Debris Disk Temperature and Stellar Type: \\ Implications for the Formation and Evolution of Wide-Orbit Planets}

\author{Nicholas P. Ballering\altaffilmark{1},
George H. Rieke\altaffilmark{1},
Kate Y. L. Su\altaffilmark{1},
Edward Montiel\altaffilmark{1,2}}

\affil{$^1$ Steward Observatory, University of Arizona, 933 North Cherry Avenue, Tucson, AZ 85721, USA}
\affil{$^2$ Department of Physics \& Astronomy, Louisiana State University, Baton Rouge, LA 70803, USA}

\begin{abstract}
Cold debris disks trace the limits of planet formation or migration in the outer regions of planetary systems, and thus have the potential to answer many of the outstanding questions in wide-orbit planet formation and evolution. We characterized the infrared excess spectral energy distributions of \NumColdComps{} cold debris disks around \NumTargs{} main-sequence stars observed by both {\it Spitzer} IRS and MIPS. We found a trend between the temperature of the inner edges of cold debris disks and the stellar type of the stars they orbit. This argues against the importance of strictly temperature-dependent processes (e.g. non-water ice lines) in setting the dimensions of cold debris disks. Also, we found no evidence that delayed stirring causes the trend. The trend may result from outward planet migration that traces the extent of the primordial protoplanetary disk, or it may result from planet formation that halts at an orbital radius limited by the efficiency of core accretion.
\end{abstract}


\section{INTRODUCTION}
\label{sec:introduction}

From the sizes of protoplanetary disks, we expect planets to form out to tens of AU. The many giant planets found on small orbits by radial velocity and transit techniques are believed to have formed beyond the water ice lines of their stars and migrated inward. Inward migration of massive planets can have a profound (and usually destructive) effect on the smaller objects in a planetary system, such as Earth-sized planets within the habitable zones. It is therefore important to determine how many systems retain their giant planets on wide orbits. Direct detection of these planets is limited to the most massive examples. Indirect detection through the use of debris disks addresses this limitation.
 
A debris disk consists of the circumstellar solid material that remains after the protoplanetary disk gas has dispersed and giant planets might have formed. Although most of the mass in a debris disk is harbored by the parent bodies (planetesimals), dust generated in their collisions accounts for the majority of the disk's surface area, and observations of debris disks via their thermal emission or reflected stellar radiation trace this dust. While tens of debris disks have been spatially resolved, the majority are detected only as an infrared excess in the spectral energy distribution (SED) of their star. Typical debris disk temperatures are tens to a few hundred kelvins, emitting as modified blackbodies that peak in the mid to far infrared. These wavelengths are well-suited for study with the {\it Spitzer Space Telescope} \citep{werner2004} Infrared Spectrograph (IRS; \citealp{houck2004}) and Multiband Imaging Photometer for {\it Spitzer} (MIPS; \citealp{rieke2004}). For a recent review of debris disks, see \citet{wyatt2008}.

Debris disks often appear constrained to one or two discrete rings. This is evident from images of resolved disks such as Fomalhaut \citep{kalas2008,boley2012}, and from SEDs of unresolved debris disks that are fit well by one or two blackbody functions, corresponding to dust at one or two distinct radial locations. Of the 28 disks without strong emission features presented by \citet{chen2006}, all but one were fit better with a blackbody function than with a continuous disk model. The exception was HR8799, which was later determined to be best modelled  by two blackbodies \citep{chen2009,su2009}. \citet{morales2009} originally fit some excess SEDs with a power law, representing a continuous radial distribution of dust; however, \citet{morales2011} argued that these power law fits require the optical depth of the disk to increase with orbital radius (which is theoretically implausible), and they found that these targets can be fit well by two blackbodies instead. These ``warm" and ``cold" debris disk components may be analogous to the asteroid belt and Kuiper belt in the Solar System.

Why do rings form, and what sets their location? Ice lines are one possibility. During the protoplanetary disk phase, a radial pressure gradient in the gas partially counteracts the gravitational force on the gas, allowing it to rotate at a sub-Keplerian velocity. Solid particles orbit at Keplerian rates, and thus experience a head wind that slows their rotation and makes them spiral inwards. When these solids reach the ice line, the volatile component sublimates, producing a local pressure increase that counteracts the overall pressure gradient. This creates a zone where the particles can settle without a headwind. Thus, there is a tendency to have a planetesimal belt near the ice line, and this belt can then produce grains that assume a specific temperature. \citet{morales2011} found similar warm disk temperatures (190 K) around stars of different stellar types. This can potentially be explained by the presence of the water ice line at 150-170 K. \citet{dodsonrobinson2009B} argue that ice lines of other species (e.g. CO, $\text{CH}_4$, $\text{N}_2$) at lower temperatures may play an important role in planet formation in the outer Solar System. If ice lines are also responsible for setting the location of the cold components, we would expect these components to have similar temperatures over a range of stellar types.
 
Planets may also cause the discrete ring structures. Planetesimals will be scattered away once a planet is massive enough to dominate the gravity in its vicinity. Observations support the expected relation between planets and disk structure: the four giant planets imaged around HR 8799 \citep{marois2010} are located in the gap between the warm and cold debris disk components \citep{su2009}; the imaged planet orbiting $\beta$ Pic appears to sculpt the inner edge of its debris disk \citep{lagrange2010}; the well-resolved cold debris ring around Fomalhaut is likely confined by planets \citep{kalas2008,boley2012}; and in the Solar System, Neptune sculpts the inner edge of the Kuiper belt \citep{liou1999}.
 
If the locations of debris disks are set by planets, we can use cold debris disks to investigate wide-orbit planets, which are not easily studied by other means. Of the over 850 confirmed exoplanets, 35 have orbits larger than 5 AU, and only 17 have orbits larger than 10 AU (NASA Exoplanet Archive\footnote{http://exoplanetarchive.ipac.caltech.edu}). This is likely due to the observational biases of the radial velocity and transit detection techniques. Direct imaging can detect wide-orbit planets, but the current technology is only sensitive to very massive planets around young, nearby stars. Uranus and Neptune, for instance, could not be directly detected from outside the Solar System.
 
The planet formation processes beyond 5-10 AU are poorly understood. For example, did the planets around HR 8799 form at their current locations or did they (or at least their cores) form on smaller orbits, then migrate outwards via scattering interactions with other planets or planetesimals? Cold debris disks provide an observational test of these planet formation and migration theories. Planet formation by core accretion becomes less efficient farther from the star due to a radial increase in the dynamical timescale and decline in the surface density of solids in the protoplanetary disk \citep{mordasini2008,dodsonrobinson2009}. Therefore, the inner edge of a cold debris disk may represent the outer limit of efficient core accretion \citep{kennedy2010,dodsonrobinson2011}. A planet migrating outwards into a debris disk will push the inner edge of the disk outward as well. Uranus and Neptune may have formed on smaller orbits and migrated into the Kuiper belt \citep{tsiganis2005}. In this alternate scenario, the inner edge of a cold debris disk may represent the limits of outward migration.

In this paper, we focus on {\it Spitzer} measurements of cold debris disks, showing how their temperatures vary with the temperature of their central star and what this implies about planet formation and migration on wide-orbits. First, we describe the selection of our target stars (\S \ref{sec:targetselection}). Then, we outline our photometric (\S \ref{sec:photometry}) and IRS (\S \ref{sec:irsdatareduction}) data acquisition/reduction. Next, we detail our modeling of the stellar photosphere SED (\S \ref{sec:photospheremodel}), our derivation of the infrared excess, and our fitting of blackbodies to the excess (\S \ref{sec:blackbodyfitting}). Finally, we analyze the results (\S \ref{sec:results}) and discuss the implications for wide-orbit planet formation and migration (\S \ref{sec:discussion}), before offering a summary and concluding remarks (\S \ref{sec:conclusions}).

\section{METHODS}
\label{sec:methods}

\subsection{Target Selection}
\label{sec:targetselection}

We searched the {\it Spitzer} observers log for main-sequence stars that were observed with the IRS Long Low (LL) module (both orders)  in staring mode and with MIPS at 24 $\micron$ and 70 $\micron$, and we accumulated a sample of \NumTargs{} targets. The stellar properties of our target list are summarized in Table \ref{table:stellarproperties}. After reducing and analyzing the data, we refined the sample as described in \S \ref{sec:blackbodyfitting}.  This yielded \NumExcess{} stars with significant excess (of which \NumColdComps{} had cold components), which are detailed in Table \ref{table:excessdata}.

It is important to note that our sample comprised stars from a variety of {\it Spitzer} observational programs, each having targets selected in a different manner (e.g. nearby stars, stars with previously detected infrared excess, stars with RV-detected planets, etc.). Hence, our sample was not selected to be statistically representative in any one sense; rather, it was designed to include as many relevant targets as possible.

We calculated the stellar temperature, $T_\star$, from the star's V-$K_s$ color using tabulated values for main-sequence stars (or interpolations between these values) from \citet{cox2000}, and we estimated a 200 K one-sigma uncertainty on these values.  $T_\star$ is listed for targets with significant excess in Table \ref{table:excessdata}. Over the range of spectral types and metallicities of our sample of stars, V-$K_s$ is an accurate indicator of stellar effective temperature, with a peak-to-peak scatter of less than $\pm$ 1.5\% and weak metallicity dependence \citep{masana2006}. We use $T_\star$ to parametrize stellar type, and this degree of accuracy is sufficient for our purpose. Some of the uncertainty in $T_\star$ reflects the effect of interstellar reddening on V-$K_s$, although the great majority of stars in this sample are within the Local Bubble and therefore should not be strongly reddened (of the \NumColdComps{} targets for which we detect cold components, only \ColdsBeyondHund{} are beyond 100 pc, \ColdsBeyondHundTwenty{} are beyond 120 pc, and \ColdsBeyondHundFifty{} are beyond 150 pc). 

Ages for these stars were estimated from a combination of chromospheric activity measurements, x-ray emission, placement on the HR diagram, surface gravity, membership in clusters and associations, and gyrochronology  collected from the literature. \citet{sierchio2013} describe the intercomparison of these methods and how they are applied, and the quoted ages are on the scale calibrated by \citet{mamajek2008}.  The ages are listed in Table \ref{table:stellarproperties}, along with references for the measurements used to derive the ages and a quality flag for the age accuracy (ranging from 0 if no age could be determined to 3 if there were three or more age measurements with good agreement).

\begin{deluxetable*}{cccccccccc}
\tabletypesize{\scriptsize}
\tablewidth{0pt}
\tablecolumns{11}
\tablecaption{Target Properties \label{table:stellarproperties}}
\tablehead{\colhead{HIP} & \colhead{HD} & \colhead{HR} & \colhead{Spectral} & \colhead{Distance} & \colhead{Age} & \colhead{Age} & \colhead{Age} & \colhead{IRS} & \colhead{Excess} \\ \colhead{} & \colhead{} & \colhead{} & \colhead{Type} & \colhead{(pc)} & \colhead{(Gyr)} & \colhead{Quality} & \colhead{References} & \colhead{AOR} & \colhead{Verdict}}
\startdata
HIP110 & HD224873 & \nodata & G5 & 49.6 & 0.42 & 2 & 2,20 & 5346816 & Not Included \\
HIP345 & HD225200 & HR9102 & A0V & 124.8 & 0.1 & 2 & 31 & 12720128 & Warm + Cold \\
HIP394 & HD225239 & HR9107 & G2V & 39.2 & 0.25 & 1 & 2 & 4028672 & Not Included \\
HIP490 & HD105 & \nodata & G0V & 39.4 & 0.17 & 3 & 2,3,4,5,6,9,12 & 5295616 & Cold \\
HIP522 & HD142 & HR6 & F7V & 25.7 & 4.8 & 2 & 3,17,27 & 22296064 & Not Included \\
HIP544 & HD166 & HR8 & K0Ve & 13.7 & 0.24 & 3 & 1,2,6,15,23 & 12717824 & Warm + Cold \\
HIP560 & HD203 & HR9 & F3Vn & 39.4 & 0.01 & 2 & 38 & 14983424 & Cold \\
HIP682 & HD377 & \nodata & G2V & 39.1 & 0.22 & 3 & 2,4,9,12,20,32 & 5268736 & Warm + Cold \\
HIP910 & HD693 & HR33 & F8Vfe-08H-05 & 18.7 & 3 & 3 & 3,4,23,24 & 14494720 & Not Included \\
& & & & & & & & 4008448 & \\
HIP919 & HD691 & \nodata & K0V & 34.2 & 0.24 & 3 & 2,9,12,14,18,20,39 & 5345280 & Not Included \\
HIP1031 & HD870 & \nodata & K0V & 20.2 & 2.3 & 2 & 3,5 & 16026112 & Cold \\
HIP1134 & HD984 & \nodata & F5 & 47.1 & 0.25 & 3 & 2,4,9,20 & 5271808 & Not Included \\
HIP1292 & HD1237 & \nodata & G8.5Vk: & 17.5 & 0.3 & 3 & 2,3,4,5,6 & 22290176 & Not Included \\
HIP1368 & \nodata & \nodata & K7 & 15 & 0.9 & 3 & 1,2,39 & 25674240 & Cold \\
HIP1473 & HD1404 & HR68 & A2V & 41.3 & 0.45 & 1 & 7 & 14160384 & Warm \\
\enddata
\tablecomments{Table \ref{table:stellarproperties} is published in its entirety in the electronic edition of the Astrophysical Journal. A portion is shown here for guidance regarding its form and content.}
\tablerefs{
(1) \citet{duncan1991};
(2) Rosat All Sky Survey \citep{voges1999,voges2000};
(3) \citet{gray2006};
(4) \citet{schroder2009};
(5) \citet{henry1996};
(6) \citet{rochapinto1998};
(7) \citet{vican2012} -- isochrone ages;
(8) \citet{schmitt2004};
(9) \citet{wright2004};
(10) \citet{katsova2011};
(11) \citet{martinez-arnaiz2010};
(12) \citet{isaacson2010};
(13) \citet{vican2012} -- gyro ages;
(14) \citet{barnes2007};
(15) \citet{gray2003};
(16) $v~\sin(i)$;
(17) \citet{jenkins2006};
(18) \citet{montes2001};
(19) \citet{vican2012} -- X-ray;
(20) \citet{white2007};
(21) $\log(g)$;
(22) \citet{lachaume1999};
(23) \citet{buccino2008};
(24) \citet{sierchio2013} -- HR diagram position;
(25) \citet{paunzen1997};
(26) \citet{nakajima2012};
(27) \citet{jenkins2011};
(28) \citet{mamajek2008};
(29) \citet{barrado1998};
(30) \citet{rhee2007};
(31) \citet{su2006};
(32) \citet{pace2013};
(33) \citet{feltzing2001};
(34) \citet{tetzlaff2010};
(35) \citet{rizzuto2011};
(36) \citet{hoogerwerf2000};
(37) \citet{plavchan2009};
(38) \citet{zuckerman2004};
(39) \citet{herrero2012};
(40) \citet{murgas2013};
(41) \citet{karatas2005};
(42) \citet{mishenina2012};
(43) \citet{takeda2007};
(44) \citet{ng1998};
(45) \citet{baines2012}.
}
\end{deluxetable*}

\subsection{Photometry}
\label{sec:photometry}

Although we built our target list around the {\it Spitzer} IRS spectra, we supported these spectra in our analysis with a suite of photometric data. The properties of the photometric systems we used are summarized in Table \ref{table:photometrybands} and the photometric data for the targets with significant excess (see \S \ref{sec:blackbodyfitting}) are given in Table \ref{table:excessdata}.

MIPS photometry at 24 $\micron$ provided an important calibration reference for the IRS data (\S \ref{sec:irsdatareduction}), while 70 $\micron$ photometry provided a crucial constraint on the temperature of cold debris disks (\S \ref{sec:blackbodyfitting}). We used our in-house debris disk pipeline to reduce and extract photometry for the MIPS data as part of the effort to preserve the legacy of {\it Spitzer} measurements on debris disk studies (\citealp{su2010}; K. Su et al. 2013, in prep). Basic reduction (up to the post-BCD mosaics) and calibrations of the MIPS data follow the descriptions by \citet{engelbracht2007} and \citet{gordon2007}. In addition, the extraction of the 24 $\micron$ photometry was briefly described in \citet{urban2012}, where the source position at 24 $\micron$ is determined by a combination of PSF fitting and 2D Gaussian fitting. Both PSF fitting and aperture photometry\footnote{The 24 $\micron$ aperture photometry used an aperture radius of 6.255$\arcsec$, a sky annulus of 19.92-29.88$\arcsec$, and an aperture correction factor of 1.6994.} were performed, and we preferentially used the aperture photometry at 24 $\micron$ because this was used to calibrate the photosphere model at 24 $\micron$, as described in \S \ref{sec:photospheremodel} and the Appendix (the PSF and aperture photometry agree to within a few percent). We then used the 24 $\micron$ source position to perform PSF fitting for data at 70 $\micron$ by minimizing the residual signal at the source position (for details, see K. Su et al. 2013, in prep). For faint sources located in areas with structured background, the resultant 70 $\micron$ photometry can be negative, which reflects non-detection. The 1$\sigma$ photometry uncertainty (listed in Table \ref{table:excessdata}) includes the pixel-to-pixel variation near the source of interest, and detector repeatability (1\% and 5\% of the source flux at 24 and 70 $\micron$, respectively). MIPS photometry data for all of our targets can be found in \citet{gaspar2013}, \citet{sierchio2013}, and K. Su et al. (2013, in prep).

V, J, H, and $K_s$ photometry were used to model the stellar photosphere SED (\S \ref{sec:photospheremodel}) and to estimate $T_\star$ (\S \ref{sec:targetselection}). We obtained {\it Hipparcos} V band \citep{ESA1997} and 2MASS J, H, and K band photometry \citep{cutri2003} from the VizieR online database. Many of the targets in the sample are nearby stars and are severely saturated in the 2MASS data. To overcome this, we used heritage aperture photometry, which we transformed to match the 2MASS system. When both 2MASS and transformed heritage photometry of high quality were available, we averaged them.  The references for the heritage photometry are in Table \ref{table:excessdata}.

IRAC photometry (3.6 $\micron$) was also used to model the stellar photosphere SED (\S \ref{sec:photospheremodel}). These data were taken in {\it Spitzer} cycle 7 (PID 70076, PI: Su). All the IRAC data were taken in subarray mode with four dithered positions to avoid saturation. We used the Basic Calibrated Data (BCD) products provided by the {\it Spitzer} Science Center (pipeline version S18.18), and performed the necessary steps (pixel solid angle correction and pixel phase correction; K. Su et al. 2013, in prep) to extract the photometry. Aperture photometry was used for each individual data frame (64 frames per dithered position); and the final quoted flux is the median value of all measurements per star.

WISE data \citep{cutri2012} were obtained from VizieR, but are not included in Table \ref{table:excessdata}. Because WISE photometry is less accurate than {\it Spitzer} photometry, we did not use it quantitatively. Instead, we used it as a qualitative confirmation of our {\it Spitzer} data.

\subsection{IRS Data Reduction}
\label{sec:irsdatareduction}

Our IRS reduction started with the Level 1 BCD products, downloaded from the {\it Spitzer} Heritage Archive. In addition to IRS LL data, we also reduced Short Low (SL) module data, when available, and included them in our analysis. The Astronomical Observation Requests (AORs) used for each target are listed in Table \ref{table:stellarproperties}. The basic reduction was performed using the Spectroscopic Modeling Analysis and Reduction Tool (SMART) software package \citep{higdon2004}, scripted into a series of automated routines. For each spectral order of each IRS AOR, three files (2D spectra, uncertainty, and mask) were combined into a single 3-plane file. Bad and rogue pixels were removed by the routine IRSCLEAN with the clean parameter set to 4096 (pixels with this value or higher were included in the clean). Next, when available, multiple Data Collection Events (DCEs) for the same nod position were combined, and then the background was removed from each 2D spectrum by subtraction of the opposite nod. The 2D spectra were then converted into 1D spectra using SMART's optimal 2 nod extraction \citep{lebouteiller2010}. The 1D spectra from both nods were combined. The result was a wavelength, flux, and uncertainty vector of each spectral order for each AOR. The ``bonus" third order data were not used. The remainder of the data reduction and analysis was performed with the MATLAB software package. 

IRS data at wavelengths near the order edges often were of poor quality, so data longward of 38 $\micron$ and shortward of 20.75 $\micron$ were discarded from the LL first order, data shortward of 14.3 $\micron$ were discarded from the LL second order, data longward of 14.7 $\micron$ and shortward of 7.55 $\micron$ were discarded from the SL first order, and data shortward of 5.25 $\micron$ were discarded from the SL second order. Systematic offsets in flux existed between IRS orders, which we fixed by applying a multiplicative correction factor to the LL first order spectrum and to both SL order spectra (if they existed), in order to align them with the LL second order flux.  Initially, we determined the order correction factors using an automated routine, but due to the variety of shapes of the spectra and the presence of outlying data points, we found that fine-tuning these factors by eye was more reliable. The data from all available orders and modules were then combined into a single spectrum.

Next, we cut outlying data points from the spectrum in an iterative process. The spectrum was fit with a polynomial and the standard deviation of residual flux values around the fit was calculated. Points lying more than three standard deviations from the fit were discarded. This process was iterated six times. Because the scatter of the data generally increased towards longer wavelengths, we applied this process separately to the short and long ends of our spectrum; if SL data were available, the two sections were divided at 14 $\micron$ and each section fit with a fourth degree polynomial, whereas if no SL data were available, the sections were divided at 25 $\micron$ and each section was fit with a second degree polynomial.  This procedure, on average, cut 10 data points from each spectrum, with the first iteration typically cutting five points, and the sixth iteration only cutting zero, one, or two points (nearly 90\% of spectra had no points cut in the sixth iteration). Before cutting, the spectra had 181, 296, or 373 points, depending on the available orders. After the outliers were cut, the spectrum was smoothed by binning to a wavelength resolution of 0.7 $\micron$.

The absolute calibration of IRS data is only accurate to $\sim$10\%, whereas the MIPS calibration is accurate to $\sim$2\% \citep{engelbracht2007}. To reduce systematic errors in our IRS data, we multiplicatively scaled our spectrum to be consistent with the measured MIPS 24 $\micron$ flux density for each target. The IRS spectra were calibrated as point sources, so if a source was slightly extended, it would have an incorrect slit-loss correction. The stellar photospheres (point-like) are dominant in the IRS spectra for the majority of sources in the sample; therefore, this has no significant impact on the result. If the MIPS photometry for a source was contaminated by background emission, this contamination would be passed to the IRS spectrum of the source through this scaling (the IRS data may or may not have picked up the contamination, depending on the slit orientation). Nevertheless, an additional multiplicative factor was later applied to the IRS data (see \S \ref{sec:photospheremodel}), which corrected any lingering errors from this process.

Some targets were observed with more than one IRS AOR. We combined the spectra from these AORs, interleaving their data points, then re-smoothing to a spectral resolution of 0.7 $\micron$. 

\subsection{Photosphere Model}
\label{sec:photospheremodel}

Although the spectrum of a main-sequence star peaks in the visible or near-IR, the stellar photosphere can still be the dominant source of flux density in the mid-IR. Thus, to extract the thermal emission of the debris disk (the infrared excess), an accurate model spectrum of the stellar photosphere must be generated and subtracted from the data.

 It is common to use photosphere spectra from detailed numerical models of stellar structures, such as KURUCZ/ATLAS9 \citep{castelli2004}. However, these models are not well-tested in the mid- and far-infrared. \citet{sinclair2010} compared various families of models (KURUCZ/ATLAS9, MARCS, and NEXTGEN/PHOENIX) and found that the choice of model family can affect whether a star is determined to have an infrared excess or not.

We opted for a simpler model, a Rayleigh-Jeans relation given by $F_{\nu\,\star}(\lambda)=\text{RJ}/\lambda^2$, where RJ is an amplitude scale factor. This model is appropriate for wavelengths beyond 5 $\mu$m because stars in this study, mostly A through K dwarfs, have virtually no spectral features in the mid-IR, behaving as blackbodies in the Rayleigh-Jeans regime.  For example, in the range of most interest for this study, the departure of a reference A0 star spectrum \citep{rieke2008} from a Rayleigh-Jeans fit between 15 and 30 $\micron$ is no more than $\pm$ 0.3\%, and the ATLAS9 \citep{castelli2013} solar model follows Rayleigh-Jeans behavior to within $\pm$ 0.24\% from 20 to 40 $\micron$. However, the theoretical spectra show minor systematic differences from the observations \citep{rieke2008}, so use of them to improve the knowledge of the SEDs would be risky.

Modeling the photosphere thus came down to finding the appropriate scale factor (RJ) for each star in our sample. We used a set of color relations (derived from a sample of stars known to have no IR excess) to predict the MIPS 24 $\micron$ magnitude of the photosphere, [24], from measured V, J, H, $K_s$, and IRAC magnitudes -- bands where  IR excesses due to very close-in dust are very rare. The details of these relations are described in the Appendix. This procedure estimates the 24 $\mu$m photospheric outputs to about 2\% rms, based solely on data at wavelengths short of 4 $\mu$m. We then converted [24] to flux density units and used it to find the scaling factor for the photosphere model, which is given in Table \ref{table:excessdata} for targets with significant excess (see \S \ref{sec:blackbodyfitting}).

We finally applied a small multiplicative factor to the IRS data  if it clearly looked slightly offset from the photosphere model. Then, we subtracted the photosphere model from the IRS spectra and MIPS 70 $\micron$ point to obtain the infrared excess, $F_{\nu\,\text{excess}}\left(\lambda\right) = F_\nu\left(\lambda\right)-F_{\nu\,\star}(\lambda)$. The photosphere model was assumed to have an uncertainty of 2\%, thus the uncertainty in the excess flux density was $$\sigma_\text{excess}\left(\lambda\right) = \sqrt{\sigma\left(\lambda\right)^2 + \left[0.02F_{\nu\,\star}(\lambda)\right]^2}.$$

\subsection{Black Body Fitting}
\label{sec:blackbodyfitting}

\begin{figure}
\begin{centering}
\includegraphics[scale=.40]{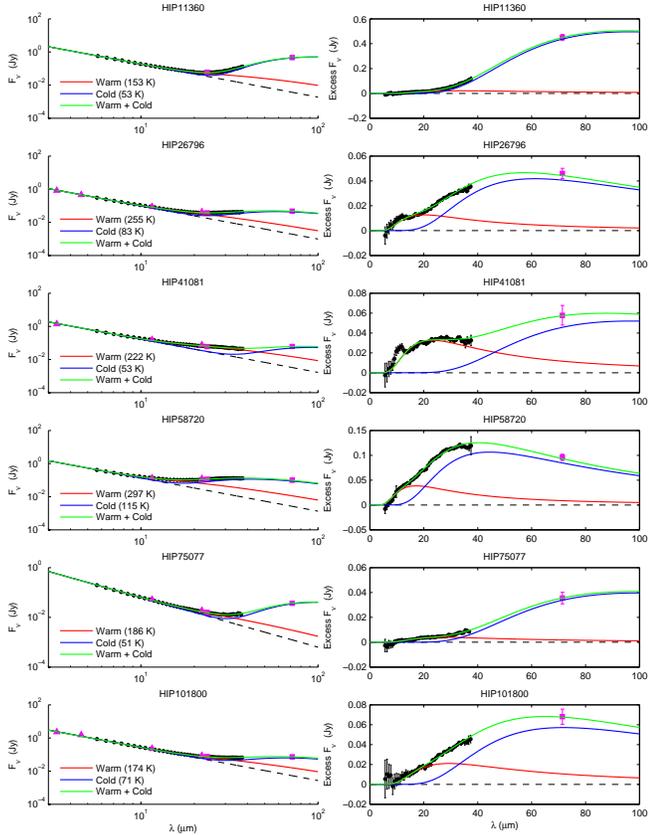}
\caption{A gallery of six targets from our sample found to have two components. The left panels include the photosphere, while the right panels show the excess above the photosphere. IRS data are black points, MIPS data are magenta squares, and WISE data are magenta triangles. The black dashed line is the photosphere, the red and blue lines are the warm and cold components of the model, respectively, and the green line is the total model. Note that error bars are omitted from the left panels for clarity. WISE and MIPS 24 $\micron$ data are omitted from the right panels, as these points were not used to constrain the fits.}
\label{fig:twocompgallery}
\end{centering}
\end{figure}

To interpret the excesses in a general way, we needed to assign them fiducial temperatures. The emission of disk grains is a function of their optical constants; however, as discussed in \S \ref{sec:introduction}, debris disks radiate like blackbodies at one or two temperatures. The simplest possible description of the emission is in terms of these temperatures.  In this section, we describe how we determined if each target had significant excess and whether the excess was best described by one or two blackbodies.

We fit $F_{\nu\,\text{excess}}\left(\lambda\right)$ with a one component blackbody model 
\begin{equation}
\label{eq:1bbmodel}
F_{\nu\,\text{model1}}(\lambda)=c_\text{one}F_{\nu\,\text{BB}}(\lambda,T_\text{one})
\end{equation}
where $F_{\nu\,\text{BB}}(\lambda,T)$ is the Planck Function. The best-fit parameters, $c_\text{one}$ and $T_\text{one}$ were found by minimizing the reduced chi-squared,
\begin{align}
\label{eq:reducedchisq}
\chi_{\nu}^2 = &\frac{1}{\nu} \Bigg\{ \sum_i^N \frac {\left[F_{\nu\,\text{excess}}\left(\lambda_i\right)-F_{\nu\,\text{model}}\left(\lambda_i\right)\right]^2} {\sigma_\text{excess}\left(\lambda_i\right)^2} \nonumber\\
&+ 28\frac {\left[F_{\nu\,\text{excess}}\left(70 \micron\right)-F_{\nu\,\text{model}}\left(70 \micron\right)\right]^2} {\sigma_\text{excess}\left(70 \micron\right)^2} \Bigg\},
\end{align}
where $\nu = N+28-n-1$, N is the number of data points in the IRS excess data, and n is the number of free parameters in the fit (n=2). The MIPS 70 $\micron$ data point was weighted in the fit as 28 IRS data points, equivalent to the number of IRS data points (0.7 $\micron$ resolution) that would fit inside the equivalent width of the MIPS 70 $\micron$ spectral response function (19.65 $\micron$). The fit was performed using MATLAB's lsqcurvefit algorithm. $T_\text{one}$ was constrained to between 0 and 500 K.  

While $c_\text{one}$ represents the amplitude of the debris disk emission, a more useful measure of a debris disk's brightness is the fractional excess, $f\equiv L_\text{excess}/L_\star$. We calculated $f$ according to
\begin{equation}
\label{eq:fwyatt2}
f=\left(\frac{F_{\nu\,\text{excess max}}}{F_{\nu\,\star\,\text{max}}}\right) \left(\frac{\lambda_{\star\,\text{max}}}{\lambda_\text{excess max}}\right),
\end{equation}
from Equation 2 of \citet{wyatt2008}. $F_{\nu\,\text{excess max}}$ and $F_{\nu\,\star\,\text{max}}$ are the peak flux density values of the disk emission and stellar photosphere emission, respectively, which occur at wavelengths $\lambda_\text{excess max}$ and $\lambda_{\star\,\text{max}}$. While the peak flux densities for the disk components were easily calculated from our best fit model, the Rayleigh-Jeans stellar photosphere model had no maximum. To overcome this, we created a blackbody function using the temperature of the star from Table \ref{table:stellarproperties}, and scaled it to match the flux density of our Rayleigh-Jeans model at 24 $\micron$. From this representation of the photosphere, we found the maximum flux density, and calculated $f$.

\begin{figure}
\begin{centering}
\includegraphics[scale=.40]{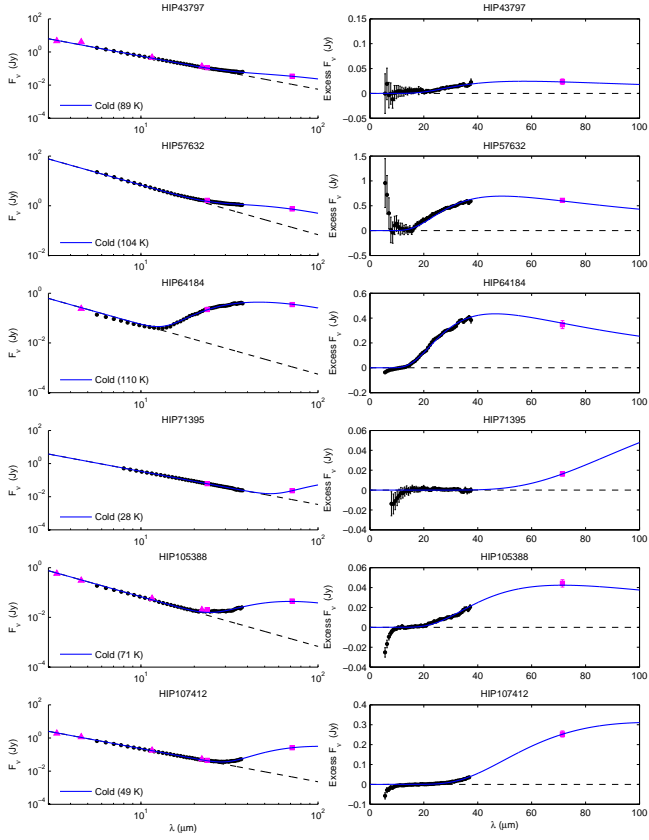}
\caption{A gallery of six targets from our sample found to have only a cold component. The left panels include the photosphere, while the right panels show the excess above the photosphere. IRS data are black points, MIPS data are magenta squares, and WISE data are magenta triangles. The black dashed line is the photosphere and the blue line is the model cold component. Note that error bars are omitted from the left panels for clarity. WISE and MIPS 24 $\micron$ data are omitted from the right panels, as these points were not used to constrain the fits.}
\label{fig:coldcompgallery}
\end{centering}
\end{figure}

With our best fits in hand, we refined our sample to only those targets with statistically significant excess. First, we inspected all fits by eye and discarded targets with poor quality data that resulted in clearly unrealistic fits. Second, we discarded targets with excess that was too faint, $f < 10^{-5}$. Third, we inspected the fits and identified all targets whose excess relied solely on the MIPS 70 $\micron$ data point (i.e. there was no excess in the IRS data). For these cases, we required this MIPS point to represent a significant excess, so we discarded targets where $F_{\nu\,\text{excess}}(70 \micron)/\sigma_\text{excess}\left(70\micron\right)<3$ or where $F_{\nu\,\text{excess}}(70 \micron)/F_{\nu\,\star}(70 \micron) < 1$. This process resulted in \NumNoExcess{} of the original \NumTargs{} stars having no significant excess.
 
Of the remaining \NumExcess{} targets with significant excess, we next determined whether the excess consisted of one or two components. To do this, we fit the excess SED of each target with a model consisting of the sum of two blackbodies,
\begin{align}
\label{eq:2bbmodel}
F_{\nu\,\text{model2}}(\lambda)&=c_\text{cold}F_{\nu\,\text{BB}}(\lambda,T_\text{cold}) \nonumber\\
&+ c_\text{warm}F_{\nu\,\text{BB}}(\lambda,T_\text{warm}).
\end{align}
Finding the optimal set of the four parameters $c_\text{cold}$, $T_\text{cold}$, $c_\text{warm}$, and $T_\text{warm}$ was again done by minimizing the reduced chi-squared (Equation \ref{eq:reducedchisq}), now with n=4.

\begin{figure}
\begin{centering}
\includegraphics[scale=.40]{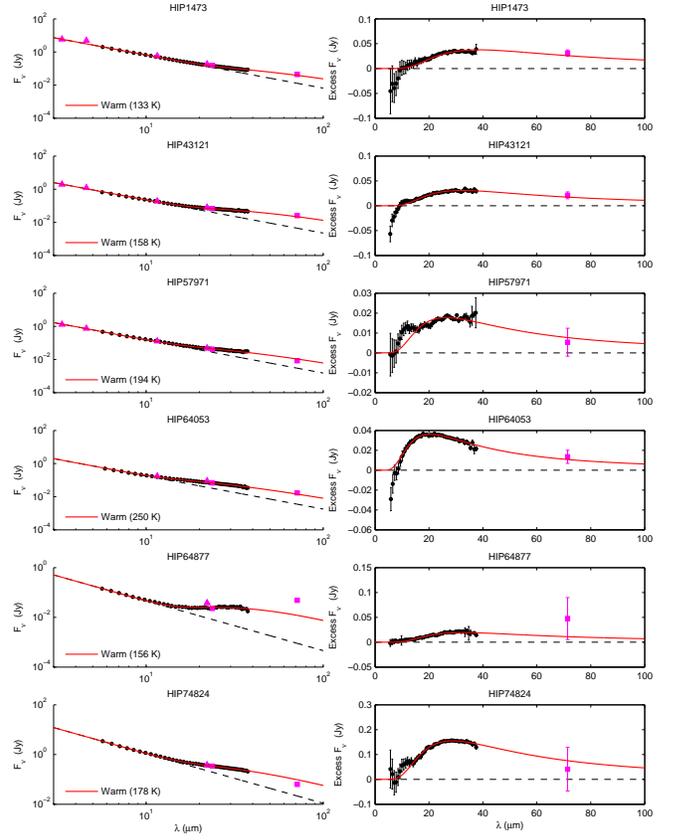}
\caption{A gallery of six targets from our sample found to have only a warm component. The left panels include the photosphere, while the right panels show the excess above the photosphere. IRS data are black points, MIPS data are magenta squares, and WISE data are magenta triangles. The black dashed line is the photosphere and the red line is the model warm component. Note that error bars are omitted from the left panels for clarity. WISE and MIPS 24 $\micron$ data are omitted from the right panels, as these points were not used to constrain the fits.}
\label{fig:warmcompgallery}
\end{centering}
\end{figure}

Our definition of ``cold" was the coldest well-detected component of the excess that was below 130 K. \citet{morales2011} found a continuous distribution of warm components above $\sim$130 K, centered around 190 K, whose temperatures were set by the water ice line. Thus, any component above 130 K was considered ``warm" for the purposes of this study. A component with a temperature of 110 K, for example, would be considered warm if another, colder temperature component was also detected, but would be considered cold if no colder component was detected.  To implement this, we fit each target once using 100 K as the division between warm and cold, and again with the division at 130 K. We then selected the better of these two cases (based on reduced chi-squared) to represent the best two-component model. For many targets, these two cases yielded identical fits. In both cases, the minimum cold temperature allowed by the fit was 0 K, and the maximum allowed warm temperature was 500 K.
 
Deciding if each target warranted a two-component model was a two-step process. First, we required that the reduced chi-squared of the two-component fit be at least three times better than that of the one-component fit (i.e. all targets where $\chi_{\nu1}^2$/$\chi_{\nu2}^2 <3$ were deemed to be better fit by a single component). Second, for the remaining targets, we identified those for which the presence of the cold component relied entirely on the MIPS 70 $\micron$ point, meaning the IRS excess was entirely fit by the warm component. For these targets to have significant cold components, we required their MIPS 70 $\micron$ excess to fall more than 3$\sigma$ above the excess predicted by the warm component. That is, if one of these targets had $$\frac{F_{\nu\,\text{excess}}(70 \micron) - c_\text{warm}F_{\nu\,\text{BB}}(70\micron,T_\text{warm})}{\sigma_\text{excess}\left(70\micron\right)}<3,$$ we concluded that it was best fit by one component. This step was analogous to the third step we performed when deciding if each target had excess or not. The targets that passed both of these criteria were deemed to have two components.

This process identified \NumOneAndCold{} single-component cold disks, \NumOneAndWarm{} single-component warm disks, and \NumTwoComps{} two-component disks. We found only one target (HIP45585) that required two warm components (both $>$130 K) to fit properly, and we discarded this target from our results as it had no cold components (it is counted in the \NumNoExcess{} targets that we discarded). We found no targets requiring two components colder than 100 K. Our methods were insensitive to possible additional cold components below $\sim$50 K. We concluded that our fitting procedure allowed us to find all cold components above this limit. Equivalently, our method reliably fit the inner edges of the cold disks.

\begin{figure}
\begin{centering}
\includegraphics[scale=.35,angle=270]{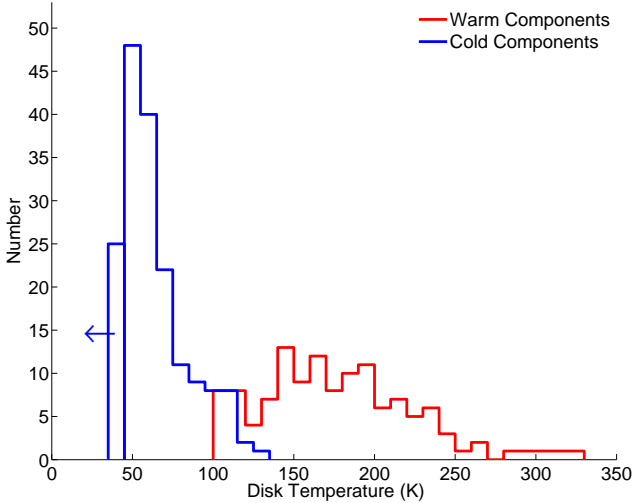}
\caption{Histograms of the warm and cold debris disk temperatures found in this sample. The bar left of 45 K contains all of the upper limit cold components.}
\label{fig:temphistogram}
\end{centering}
\end{figure}

For the two-component disks, we calculated the fractional excesses for the warm and cold components, $f_\text{warm}$ and $f_\text{cold}$, using Equation \ref{eq:fwyatt2}. One-component excesses were deemed warm or cold depending if they had temperatures above or below 130 K. The verdicts for all targets are given in Table \ref{table:stellarproperties} and the parameters of our best fits for targets with significantly-detected components are listed in Table \ref{table:excessdata}.

We estimated that the uncertainty in $T_\text{cold}$ was 10 K. This estimate was conservative; warmer cold components had more IRS data points in excess of the photosphere, so their fits were more constrained, with a temperature uncertainty of 3 to 7 K.

We were unable to accurately constrain the temperatures of very cold components, which were detected only at 70 $\mu$m. With only one data point, blackbodies at a wide range of cold temperatures could be given the amplitude necessary to match the point, so a degeneracy existed between T and $f$ in blackbody fits to the excess SED. We made this distinction at 45 K; above this temperature there was enough information in the IRS spectra to constrain the temperature. Cold temperatures above 45 K were considered true detections, while cold temperatures below 45 K were considered unconstrained and were replaced with upper limits at 45 K.  By using the 70 $\micron$ data to determine if such targets had significant excesses, we ensured that these disks were truly very cold, rather than warm but very faint. Of the \NumColdComps{} cold components in our sample, \NumUpperLims{} had temperature upper limits.

Some IRS spectra exhibit mineralogical emission features (e.g. HIP41081, HIP57971). Our model did not explicitly account for these features, but they peak sharply at $\sim$10 $\micron$ and do not resemble blackbody functions. These features did not influence our fits, except to raise the value of the minimum reduced chi-squared.

The data and best fits for a small sample of our targets are shown in Figures \ref{fig:twocompgallery}, \ref{fig:coldcompgallery}, and \ref{fig:warmcompgallery} (for targets with two components, one cold component, and one warm component, respectively). The left panels in all of these figures show the measured data with the photosphere and blackbody models, while the right panels illustrate the excess data and models with the photosphere subtracted. Histograms of the significant warm and cold debris disk temperatures are shown in Figure \ref{fig:temphistogram}.

\section{RESULTS}
\label{sec:results}

\begin{figure}
\begin{centering}
\includegraphics[scale=.35,angle=270]{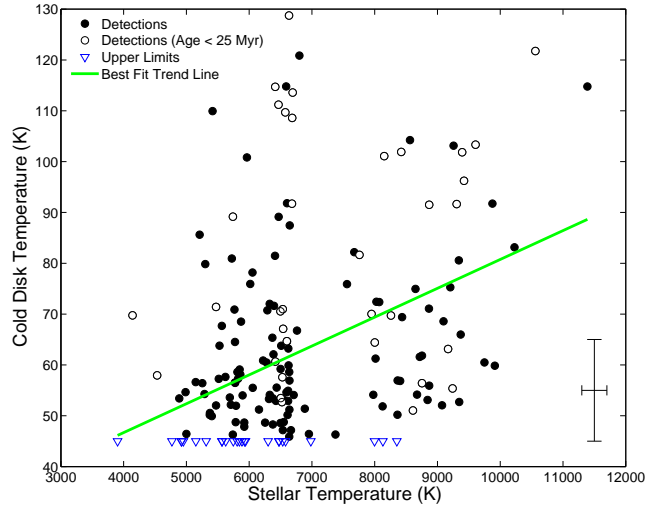}
\caption{The temperature of the cold disk component versus the temperature of the disk's host star. Black circles are well-determined disk temperatures (open circles are young systems), and blue triangles are upper limits. Although there is substantial scatter in the cold component temperatures, a correlation between cold debris disk temperature and stellar temperature is evident. Note, for example, that there are no disks colder than 50 K around stars hotter than 8500 K, in comparison with the large number of disks with temperature $<$ 45 K around cooler stars. The green line is the best fit trend to the data, determined by a Bayesian linear regression, $T_\text{cold}= \Slope \times T_\star + \Intercept$. A representative error bar is in the lower right.}
\label{fig:TdiskTstarScatter}
\end{centering}
\end{figure}

\cite{morales2011} studied the behavior of the warm disk components in detail, but their sample  (restricted to stars with ages less than 1 Gyr) had too few 70 $\micron$ detections of solar-type stars (9) to determine any trends between the cold component temperature and stellar type. To look for such a trend in our larger sample, we plot the cold component temperature versus stellar temperature in Figure \ref{fig:TdiskTstarScatter}. Well-constrained disk temperatures are black circles (open circles are young systems with age less than 25 Myr), and upper limits are downward facing blue triangles. Although there is substantial scatter in the cold component temperatures, a positive correlation between cold debris disk temperature and stellar temperature is evident.

\begin{figure}
\begin{centering}
\includegraphics[scale=.35,angle=270]{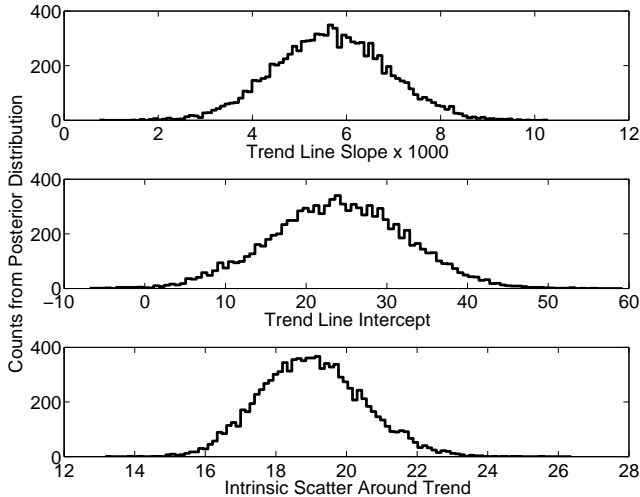}
\caption{Histograms of 10000 samples from the Bayesian posterior distributions of the $T_\text{cold}$ vs. $T_\star$ trend slope, intercept, and intrinsic scatter, generated by the linmix\_err routine.}
\label{fig:TrendPosteriors}
\end{centering}
\end{figure}

To quantify this trend, we fit a linear relation to the data by Bayesian analysis using the routine linmix\_err\footnote{http://idlastro.gsfc.nasa.gov/ftp/pro/math/linmix\_err.pro}. The routine properly handles upper-limits, and it uses uncertainties in both x and y directions (we assumed an uncertainty of 10 K in $T_\text{cold}$ and 200 K in $T_\star$). In its model, the routine also includes the normally-distributed intrinsic scatter of the data around the trend. The posterior distributions of the slope, intercept, and intrinsic scatter from the Bayesian regression are show in Figure \ref{fig:TrendPosteriors}. The slope of the linear regression was $\Slope \pm \SlopeSigma$, thus the existence of a trend between $T_\text{cold}$ and $T_\star$ is significant at level greater than 4.5$\sigma$. The 1$\sigma$ intrinsic scatter around the trend was $\Scatter \pm \ScatterSigma$ K. The best-fit trend line from the Bayesian analysis, $T_\text{cold}= \Slope \times T_\star + \Intercept$, is plotted in green in Figure \ref{fig:TdiskTstarScatter}. The choice of a linear fit is not physically motivated; its purpose is to show that -- to first order -- there is a correlation between cold debris disk temperature and stellar type.
 
A significant source of scatter around the trend results from the 10 targets with $T_\star < 7000$ K and $T_\text{cold} > 100$ K. Six of these stars (HIP560, HIP64184, HIP64995, HIP65875, HIP67497, and HIP78663) are less than 25 Myr old (most are part of the Scorpius-Centaurus Association). Although we show in \S \ref{sec:discussion} that there is no significant trend in $T_\text{cold}$ with age across the broad range of stellar ages in our sample, the excesses of these targets may be a product of their relatively young ages. They represent a period (5-50 Myr) when considerable collisional activity may still be occurring as a result of planet building \citep{kenyon2008}.

\section{DISCUSSION}
\label{sec:discussion}

We now consider what determines the temperature of cold debris disks and how a trend with stellar type might arise.

As discussed in \S 1, \citet{morales2011} found similar warm disk temperatures (190 K) around stars of different stellar types, and they attributed the effect to a particle trap at the water ice line. If ice lines of other species set the location of cold debris disks, we would expect cold disks around different type stars to have a common temperature. However, the trend of cold component temperature with stellar type (Figure \ref{fig:TdiskTstarScatter}) is inconsistent with any strictly temperature-dependent mechanism for setting the location of cold debris disks. \citet{kennedy2008} show that during the protoplanetary disk phase, the water ice line location is time dependent, and predicting it requires accounting for viscous heating in the disk, as well as the evolution of the pre-main-sequence stellar luminosity. Such considerations have not been applied to potential ice lines of species other than water, and we do not consider this level of detail here.

\begin{figure}
\begin{centering}
\includegraphics[scale=.35,angle=270]{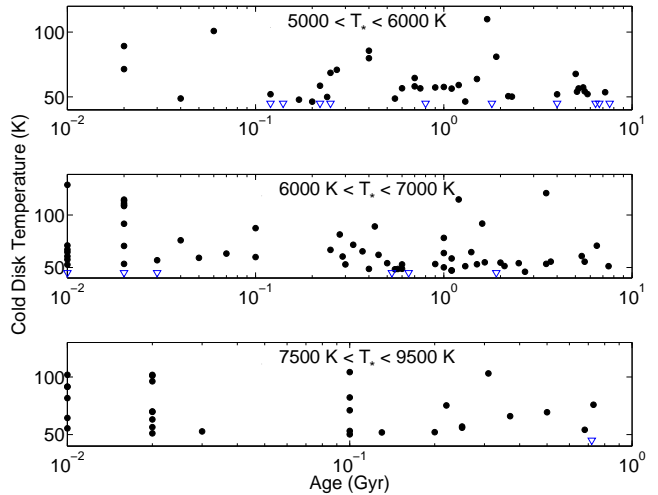}
\caption{$T_\text{cold}$ is plotted against the age of the system for targets in three $T_\star$ bins. No trend is seen with age, suggesting age does not play a confounding role in our discovered $T_\text{cold}$ vs. $T_\star$ trend. This also argues against the occurrence of delayed stirring.}
\label{fig:AgeVsTdisk}
\end{centering}
\end{figure}

Delayed stirring, in which the radial location of dust in a debris disk moves outwards with time, is a postulated mechanism for debris disk evolution  \citep{kennedy2010}. This could occur if the parent bodies are distributed in a broad ring. Particles in a debris disk collide and grind into dust faster on smaller orbits where the dynamical timescale is shorter. Thus, the location of the emitting dust moves outward with time, and, therefore, becomes cooler with time. Because late type stars have longer lifetimes than early type stars, the late type stars in our sample are generally older than the early type stars. If delayed stirring occurs, then this age bias in our sample could explain the observed trend in $T_\text{cold}$ with $T_\star$. To test this hypothesis, we plot $T_\text{cold}$ against the age of the system (if available) for targets in three $T_\star$ bins (5000 to 6000 K, 6000 to 7000 K, and 7500 to 9500 K), shown in Figure \ref{fig:AgeVsTdisk}. We see no trend in $T_\text{cold}$ with age in any bin, suggesting that delayed stirring does not produce the trend of disk temperature with stellar type.

Perhaps cold debris disks are all at roughly the same orbital distance, regardless of stellar type.  Assuming the disk is heated to its equilibrium temperature, the relation between the disk's temperature, location, and stellar type is given by  
\begin{align}
\label{eq:equilibriumtemp}
T_\text{disk} &\propto R_\text{disk}^{-1/2}L_\star^{1/4} \nonumber\\
&\propto R_\text{disk}^{-1/2}M_\star \\
&\propto R_\text{disk}^{-1/2}T_\star^2. \nonumber
\end{align}
The second and third lines of Equation \ref{eq:equilibriumtemp} are derived assuming $L_\star \propto M_\star^4$ and $L_\star \propto T_\star^8$ (valid in the roughly solar mass range). So with $R_\text{disk}$ constant, early type stars would host warmer disks. \citet{kenyon2008} perform detailed simulations of the evolution of debris disks with inner and outer edges at 30 and 150 AU, respectively, around stars with mass ranging from 1 to 3 $M_\odot$. Their models output the disk emission at 24 and 70 $\micron$, which show that the (color) temperature of these cold debris disks does increase with stellar mass, in agreement with this expectation.

But is the orbital location of cold debris disks truly constant with stellar type? If the size of cold debris disks traces the size of the original protoplanetary disks, we can use observations of the latter to address this question. \citet{andrews2010} find that protoplanetary disk mass and radius are related by
\begin{equation}
\label{eq:ppmass}
M_\text{disk} \propto R_\text{disk}^{1.6}.
\end{equation}
Furthermore, observations \citep{scholz2006} show that
\begin{equation}
\label{eq:ppandstarmass}
M_\text{disk} \propto M_\star.
\end{equation}
Combining these relations yields 
\begin{equation}
\label{eq:ppradius}
R_\text{disk} \propto M_\star^{0.63},
\end{equation}
so protoplanetary disk size does increase with earlier stellar type, albeit slowly. How quickly must the disk size increase for it to maintain a constant temperature? From Equation \ref{eq:equilibriumtemp}, 
\begin{equation}
\label{eq:equilibriumradius}
R_\text{disk}\left(T_\text{disk}=\text{const}\right) \propto M_\star^2.
\end{equation}
Although the size of disks does increase with earlier spectral type, it does so more slowly than required to maintain a constant temperature, thus disks are expected to be warmer around early type stars, consistent with our findings.  Substituting Equation \ref{eq:ppradius} into Equation \ref{eq:equilibriumtemp} reveals how the disk temperature would vary with spectral type in this case:
\begin{align}
\label{eq:migresult}
T_\text{disk} &\propto L_\star^{0.17} \nonumber\\
&\propto M_\star^{0.69} \\
&\propto T_\star^{1.38}. \nonumber
\end{align}
New results from \citet{mohanty2013} suggest that Equation \ref{eq:ppandstarmass} may hold only for $M_\star \leq 1M_\sun$, with disk mass constant or possibly even decreasing as $M_\text{disk} \propto M_\star^{-1/2}$ for higher mass stars. If this were true and translated into a flat or decreasing $R_\text{disk}$ with $M_\star$, then the disk temperature would increase even faster with spectral type than derived in Equation \ref{eq:migresult}.

Is it plausible that the inner edge of cold debris disks scales with the size of the original protoplanetary disk? Some mechanism must clear the material inside cold debris disks and set their inner edge, and planets are a common explanation. An outwardly migrating planet would set an inner edge, but migration from planet-planet scattering can be a chaotic and unpredictable phenomenon; simulations show that the planet's final location depends sensitively on the initial conditions of the system \citep{tsiganis2005}. This suggests that a trend with stellar temperature would not arise. Outward migration via scattering through a smooth disk of planetesimals would proceed in a more orderly manner, and the planet would halt its migration when the surface density of planetesimals decreased below a certain threshold, at a location that scales with the size of the original protoplanetary disk.

Planet formation by core accretion (without migration) would also clear debris inside the cold component and create its inner edge. Planetesimals would be either incorporated into the planets as they formed, or scattered away once the planets became massive enough to dominate the gravitational field in their vicinity. As mentioned in \S \ref{sec:introduction}, core accretion efficiency declines with increasing orbital radius, leaving an outer zone of planetesimals beyond the planets. The timescale for the formation of a planet with a given mass scales as 
\begin{equation}
\label{eq:formationtime}
t \propto P/\Sigma, 
\end{equation}
where P is the orbital period and $\Sigma$ is the surface density of solids. Substituting Kepler's Third Law,
\begin{equation}
\label{eq:keplerslaw}
P \propto a^{3/2}M_\star^{-1/2},
\end{equation}
($a$ is the orbital distance) and the typical protoplanetary disk structure,
\begin{equation}
\label{eq:ppsurfacedensity}
\Sigma \propto a^{-3/2}M_\star,
\end{equation}
 \citep{weidenschilling1977,kenyon2008} into Equation \ref{eq:formationtime} gives 
\begin{equation}
\label{eq:formationtime2}
t \propto a^3 M_\star^{-3/2}.
\end{equation}
After a time ($t$) planets will have formed out to a given orbital location ($a$), which sets the inner edge of the cold debris disk. So setting $t$ constant and $a=R_\text{disk}$ predicts that the size of the disk would scale with spectral type as
\begin{equation}
\label{eq:formradius}
R_\text{disk} \propto M_\star^{1/2}.
\end{equation}
By comparing this with Equation \ref{eq:equilibriumradius}, we see that in this case as well, the size of the disk grows more slowly with stellar type than required to maintain a constant equilibrium temperature, consistent with our observational results. Substituting Equation \ref{eq:formradius} into Equation \ref{eq:equilibriumtemp} shows how the disk temperature would vary with stellar type if its inner edge were set by the limits of planet formation:
\begin{align}
\label{eq:formresult}
T_\text{disk} &\propto L_\star^{3/16} \nonumber\\
&\propto M_\star^{3/4} \\
&\propto T_\star^{3/2}. \nonumber
\end{align}

\section{CONCLUSIONS}
\label{sec:conclusions}

We studied the circumstellar environment of 546 main-sequence stars via the mid infrared emission of their debris disks, as measured by the {\it Spitzer} Space Telescope. After subtracting a model of the flux expected from the stellar photosphere, we obtained an SED of the infrared excess for each target.  We found \NumExcess{} targets with significant excess: \NumOneAndCold{} with a single cold component, \NumOneAndWarm{} with a single warm component, and \NumTwoComps{} with two components. Examining the results revealed a trend between the temperature of the inner edge of the cold debris disk component and that of its host star\footnote{There is a suggestion that this trend is not well-established for systems less than 25 Myr old.}.

\begin{figure}
\begin{centering}
\includegraphics[scale=.35,angle=270]{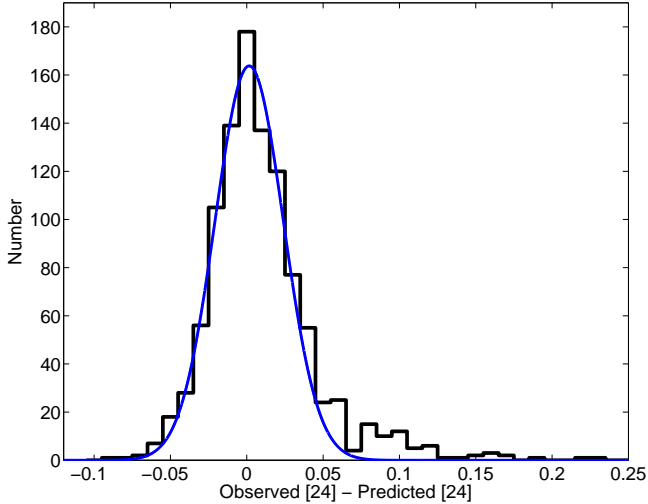}
\caption{The distribution of the difference between the observed and predicted 24 $\micron$ photosphere magnitude for a sample of 1037 stars. The blue curve shows the Gaussian fit to the distribution with $\mu=0.00158$ and $\sigma=0.0224$.}
\label{fig:24Accuracy}
\end{centering}
\end{figure}

This trend is inconsistent with theories that predict the location of cold debris disks to be strictly temperature-dependent, i.e. we rule out the dominance of ice lines in sculpting the outer regions of planetary systems. We also rule out delayed stirring as the source of this trend. The trend can potentially be explained if the outward migration of planets traces the extent of the primordial protoplanetary disk, which tends to be limited to warmer equilibrium temperatures for hotter stars. The trend can also be explained if planets form in situ out to a distance where the core accretion efficiency drops below a certain threshold, leaving a cold debris disk that is warmer around earlier type stars.   

\acknowledgments
We thank Scott Kenyon and Benjamin Bromley for helpful science discussions, and the referee for many useful comments. We also thank Vianney Lebouteiller for help in scripting the IRS reduction pipeline. This work was supported by contract 1255094 from Caltech/JPL to the University of Arizona. We used the SIMBAD database and the VizieR catalogue access tool, operated at CDS, Strasbourg, France. We also used the NASA/IPAC Infrared Science Archive, which is operated by the Jet Propulsion Laboratory, California Institute of Technology, under contract with NASA. This work is based on observations made with the {\it Spitzer} Space Telescope, which is operated by the Jet Propulsion Laboratory, California Institute of Technology under a contract with NASA. And we used data products from the Two Micron All Sky Survey, which is a joint project of the University of Massachusetts and the Infrared Processing and Analysis Center/California Institute of Technology, funded by NASA and the NSF. Finally, we used data products from the Wide-field Infrared Survey Explorer, which is a joint project of the University of California, Los Angeles, and the Jet Propulsion Laboratory/California Institute of Technology, funded by NASA.

\appendix
\section{DERIVING PHOTOSPHERE [24]}
\label{sec:derivephotosphere24}

Here we describe how we used photometry to derive a model photosphere magnitude at 24 $\micron$. The relation among V, $K_s$, and [IRAC] used to predict [24] is Equation \ref{eq:MIPS24fromIVK}, but a number of other relations were used beforehand to improve the accuracy of the result. First, we preprocessed $K_s$, H, and J photometry to remove small nonlinearities in the relations. These small corrections to the 2MASS photometry were derived by comparison with our all-sky warm mission measurements in IRAC Band 1. The preprocessing replaced $K_s$ with $K_s'$ according to
\begin{equation}
\label{eq:Kpreprocess}
K_s' = K_s - 0.0027K_s^3 + 0.0336K_s^2 - 0.1267K_s + 0.1411,
\end{equation}
H with H' according to
\begin{equation}
\label{eq:Hpreprocess}
H' = H - 0.0027K_s^3 + 0.0336K_s^2 - 0.1267K_s + 0.1411,
\end{equation}
and J with J' according to
\begin{equation}
\label{eq:Jpreprocess}
J' = J + 0.0013K_s - 0.0071.
\end{equation}

Equation \ref{eq:MIPS24fromIVK} utilizes $K_{s\,\text{SUPER}}$, a combination of measured and derived $K_s$ magnitudes. One of these, $K_{s\,1}$, was derived by solving the relation among V, $K_{s\,1}$, and J', given by
\begin{equation}
\label{eq:K1}
J'-K_{s\,1} = -0.0017x_1^4 + 0.0046x_1^3 + 0.0215x_1^2 + 0.2192x_1 - 0.0272,
\end{equation}
where $x_1=V-K_{s\,1}$. This relation is a fit to the behavior of more than 1000 main-sequence stars, as are the other relations that are given below. $K_{s\,2}$ was derived (if H was measured) by solving the relation among V, $K_{s\,2}$, and H', given by
\begin{equation}
\label{eq:K2}
H'-K_{s\,2} = -0.0011x_2^4 + 0.0108x_2^3 - 0.0327x_2^2 + 0.0739x_2 + 0.0199,
\end{equation}
where $x_2=V-K_{s\,2}$. $K_{s\,\text{SUPER}}$ was then calculated by averaging $K_s'$, $K_{s\,1}$, and $K_{s\,2}$ with relative weights of 1, 0.75, and 0.47, respectively. The weights were determined from typical 2MASS errors and by minimizing the residuals in comparing the projected photospheric levels with the 24 $\mu$m measurements. 

Equation \ref{eq:MIPS24fromIVK} also uses $[\text{IRAC}]_\text{SUPER}$, a combination of measured and derived IRAC magnitudes. The derived IRAC magnitude, $[\text{IRAC}]_1$, was calculated by solving
\begin{equation}
\label{eq:IRACderived}
[\text{IRAC}]_1 = K_{s\,\text{SUPER}} + 0.0012x^4 - 0.01448x^3 + 0.0419x^2 - 0.056x + 0.0295,
\end{equation}
where $x=V-K_{s\,\text{SUPER}}$. The measured [IRAC] and derived $[\text{IRAC}]_1$ were averaged together with relative weights of 1 and 0.5, respectively, yielding $[\text{IRAC}]_\text{SUPER}$.

Finally, [24] was calculated with
\begin{equation}
\label{eq:MIPS24fromIVK}
[24] = [\text{IRAC}]_\text{SUPER} + 0.001x^5 - 0.0097x^4 + 0.0179x^3 + 0.0305x^2 - 0.0821x + 0.0312,
\end{equation}
where, again, $x=V-K_{s\,\text{SUPER}}$. This process required V and at least one of J, H, or $K_s$ to proceed. Targets that did not have this minimum photometry available were not included in our sample, although nearly all targets in our sample had a complete suite of V, J, H, and $K_s$ photometry available (Table \ref{table:excessdata}).
 
The accuracy and precision of this method is illustrated in Figure \ref{fig:24Accuracy}, which shows the distribution of the difference between the observed and predicted [24] for over 1000 stars. A Gaussian fit to this distribution (the blue curve) has a mean of 0.00158 and a standard deviation of 0.0224.

\bibliographystyle{apj}


\clearpage
\begin{landscape}
\begin{deluxetable}{cccccccccccccccc}
\tabletypesize{\scriptsize}
\tablewidth{0pt}
\tablecolumns{14}
\tablecaption{Photometry and IR Excess Properties \label{table:excessdata}}
\tablehead{\colhead{HIP} & \colhead{V} & \colhead{J} & \colhead{H} & \colhead{$K_s$} & \colhead{Phot.} & \colhead{$T_\star$} & \colhead{IRAC} & \colhead{MIPS24} & \colhead{MIPS70} & \colhead{RJ} & \colhead{MIPS70 Excess} & \colhead{$T_\text{cold}$} & \colhead{$T_\text{warm}$} & \colhead{$f_\text{cold}$} & \colhead{$f_\text{warm}$} \\ \colhead{} & \colhead{(mag)} & \colhead{(mag)} & \colhead{(mag)} & \colhead{(mag)} & \colhead{Refs.} & \colhead{(K)} & \colhead{(Jy)} & \colhead{(mJy)} & \colhead{(mJy)} & \colhead{($\text{Jy}\times\micron^2$)} & \colhead{(mJy)} & \colhead{(K)} & \colhead{(K)} & \colhead{($\times 10^{-5}$)} & \colhead{($\times 10^{-5}$)}}
\startdata
HIP345 & 6.39 & 6.28 & 6.25 & 6.26 & \nodata & 8843 & \nodata & 35.7 $\pm$ 0.38 & 97.24 $\pm$ 5.35 & 11.9 & 94.9 $\pm$ 5.35 & 53 & 185 & 9.42 & 4.56 \\ 
HIP490 & 7.53 & 6.46 & 6.19 & 6.12 & \nodata & 5923 & \nodata & 28.39 $\pm$ 0.3 & 152.7 $\pm$ 9.71 & 14.2 & 149.9 $\pm$ 9.71 & 48 & \nodata & 42 & \nodata \\ 
HIP544 & 6.13 & 4.73 & 4.63 & 4.31 & \nodata & 5405 & \nodata & 159.1 $\pm$ 1.58 & 105.8 $\pm$ 6.32 & 72.6 & 91.57 $\pm$ 6.33 & 50 & 126 & 4.99 & 4.78 \\ 
HIP560 & 6.19 & 5.45 & 5.33 & 5.24 & \nodata & 6635 & \nodata & 117.2 $\pm$ 1.12 & 67.08 $\pm$ 7.06 & 31.6 & 60.89 $\pm$ 7.06 & 129 & \nodata & 18.4 & \nodata \\ 
HIP682 & 7.59 & 6.42 & 6.15 & 6.12 & \nodata & 5809 & \nodata & 36.49 $\pm$ 0.39 & 170.6 $\pm$ 10.62 & 14.7 & 167.7 $\pm$ 10.6 & $<$45 & 119 & 47.3 & 11.6 \\ 
HIP1031 & 7.23 & 5.85 & 5.47 & 5.38 & \nodata & 5372 & 1.9 $\pm$ 0.03 & 50.67 $\pm$ 0.51 & 22.51 $\pm$ 4.98 & 27.8 & 17.07 $\pm$ 4.99 & 50 & \nodata & 3.29 & \nodata \\ 
HIP1368 & 9 & 6.38 & 5.75 & 5.58 & \nodata & 3903 & 1.7 $\pm$ 0.03 & 46.6 $\pm$ 0.48 & 20.74 $\pm$ 5.15 & 26.5 & 15.55 $\pm$ 5.15 & $<$45 & \nodata & 9.99 & \nodata \\ 
HIP1473 & 4.52 & 4.34 & 4.42 & 4.46 & \nodata & 9164 & \nodata & 154.4 $\pm$ 1.56 & 43.82 $\pm$ 6.5 & 65.7 & 30.94 $\pm$ 6.51 & \nodata & 133 & \nodata & 1.38 \\ 
HIP1481 & 7.46 & 6.46 & 6.25 & 6.15 & \nodata & 6258 & \nodata & 34.76 $\pm$ 0.36 & -0.83 $\pm$ 2.98 & 13.7 & -3.518 $\pm$ 2.98 & \nodata & 217 & \nodata & 11.9 \\ 
HIP1499 & 6.46 & 5.33 & 5.04 & 4.89 & \nodata & 5692 & 2.95 $\pm$ 0.03 & 80.94 $\pm$ 0.81 & 61.43 $\pm$ 6.81 & 43.1 & 52.98 $\pm$ 6.81 & 54 & \nodata & 5.54 & \nodata \\ 
HIP2072 & 3.9 & 3.74 & 3.58 & 3.51 & \nodata & 8131 & \nodata & 311.2 $\pm$ 3.1 & 72.97 $\pm$ 4.43 & 143 & 44.94 $\pm$ 4.47 & $<$45 & 169 & 0.577 & 1.47 \\ 
HIP2472 & 4.77 & 4.67 & 4.77 & 4.7 & \nodata & 9096 & \nodata & 112.5 $\pm$ 1.12 & 76.98 $\pm$ 6.54 & 49.5 & 67.27 $\pm$ 6.54 & 69 & 192 & 1.47 & 1.5 \\ 
HIP2578 & 5.07 & 5.06 & 5.16 & 4.99 & \nodata & 9028 & \nodata & 232 $\pm$ 2.31 & 56.48 $\pm$ 4.13 & 35.5 & 49.52 $\pm$ 4.13 & \nodata & 194 & \nodata & 15.6 \\ 
HIP2710 & 6.91 & 6.04 & 5.85 & 5.75 & \nodata & 6473 & 1.33 $\pm$ 0.02 & 40.62 $\pm$ 0.44 & 104.7 $\pm$ 6.5 & 19.8 & 100.8 $\pm$ 6.5 & $<$45 & 110 & 16.5 & 2.94 \\ 
HIP2843 & 6.71 & 5.84 & 5.66 & 5.59 & \nodata & 6510 & 1.57 $\pm$ 0.03 & 42.13 $\pm$ 0.46 & 21.54 $\pm$ 3.76 & 23.3 & 16.98 $\pm$ 3.76 & 64 & \nodata & 2.43 & \nodata \\ 
\enddata
\tablecomments{Table \ref{table:excessdata} is published in its entirety in the electronic edition of the Astrophysical Journal. A portion is shown here for guidance regarding its form and content.}
\tablerefs{For heritage photometry used in addition to 2MASS.
(1) Obtained from SAAO (unpublished);
(2) ESO \citep{bouchet1991,vanderbliek1996};
(3) \citet{johnson1966};
(4) \citet{carter1990};
(5) \citet{glass1974};
(6) \citet{aumann1991};
(7) \citet{allen1983};
(8) \citet{groote1983};
(9) \citet{kidger2003};
(10) \citet{alonso1998}.
}
\end{deluxetable}
\clearpage
\end{landscape}

\clearpage
\begin{deluxetable}{ccc}
\tabletypesize{\scriptsize}
\tablewidth{0pt}
\tablecaption{Photometry Band Properties \label{table:photometrybands}}
\tablehead{\colhead{Band} & \colhead{$\lambda$ ($\micron$)} & \colhead{Zero Point (Jy)}}
\startdata
{\it Hipparcos} V  & 0.5423 & 3729   \\
2MASS J      & 1.241  & 1623   \\
2MASS H      & 1.6513 & 1075   \\
2MASS $K_s$      & 2.1657 & 676    \\
WISE 1       & 3.35   & 309.54 \\ 
WISE 2       & 4.60   & 171.79 \\
WISE 3       & 11.56  & 31.67  \\
WISE 4       & 22.09  & 8.36   \\
IRAC 1       & 3.6    & 269.53 \\
MIPS 24      & 23.675 & 7.17   \\
MIPS 70      & 71.42  & 0.778  \\
\enddata
\tablecomments{Properties of photometry bands that were used. J, H, and $K_s$ properties are from \citet{rieke2008}, V band properties are from \citet{holberg2006}, and WISE properties are from \citet{jarrett2011}. An IRAC 1 zero point of 283.25 Jy is typical, but we adjusted it to 269.53 to bring our newly-reduced IRAC magnitudes in line with an older IRAC reduction, from which Equation \ref{eq:MIPS24fromIVK} was derived.}
\end{deluxetable}

\end{document}